# Whence the Minkowski Momentum?

Masud Mansuripur[†] and Armis R. Zakharian[‡]

[†]College of Optical Sciences, The University of Arizona, Tucson, Arizona 85721
[‡]Corning Incorporated, Science and Technology Division, Corning, New York 14831

**Abstract**. Electromagnetic waves carry the Abraham momentum, whose density is given by $\boldsymbol{p}_{EM} = \boldsymbol{S}(\boldsymbol{r},t)/c^2$. Here $\boldsymbol{S}(\boldsymbol{r},t) = \boldsymbol{E}(\boldsymbol{r},t) \times \boldsymbol{H}(\boldsymbol{r},t)$ is the Poynting vector at point $\boldsymbol{r}$ in space and instant $t$ in time, $\boldsymbol{E}$ and $\boldsymbol{H}$ are the local electromagnetic fields, and $c$ is the speed of light in vacuum. The above statement is true irrespective of whether the waves reside in vacuum or within a ponderable medium, which medium may or may not be homogeneous, isotropic, transparent, linear, magnetic, etc. When a light pulse enters an absorbing medium, the force experienced by the medium is only partly due to the absorbed Abraham momentum. This absorbed momentum, of course, is manifested as Lorentz force (while the pulse is being extinguished within the absorber), but not all the Lorentz force experienced by the medium is attributable to the absorbed Abraham momentum. We consider an absorptive/reflective medium having the complex refractive index $n_2 + i\kappa_2$, submerged in a transparent dielectric of refractive index $n_1$, through which light must travel to reach the absorber/reflector. Depending on the impedance-mismatch between the two media, which mismatch is dependent on $n_1$, $n_2$, $\kappa_2$, either more or less light will be coupled into the absorber/reflector. The dependence of this impedance-mismatch on $n_1$ is entirely responsible for the appearance of the Minkowski momentum in certain radiation pressure experiments that involve submerged objects.

**1. Introduction**. The long-standing Abraham-Minkowski controversy [1-7] surrounding the momentum of light inside dielectric media is the consequence of a subtle defect of the classical theory of electrodynamics. Maxwell's macroscopic equations are self-consistent as well as consistent with the special theory of relativity, but they are incomplete in the sense that the energy and momentum of the electromagnetic field cannot be *derived* from these equations [8]. Over the years, correct expressions have been discovered for the field's energy density in the most general case involving the fields $\boldsymbol{E}(\boldsymbol{r},t)$, $\boldsymbol{H}(\boldsymbol{r},t)$, $\boldsymbol{D}(\boldsymbol{r},t)$, $\boldsymbol{B}(\boldsymbol{r},t)$ and their sources $\rho_{\text{free}}(\boldsymbol{r},t)$, $\boldsymbol{J}_{\text{free}}(\boldsymbol{r},t)$, $\boldsymbol{P}(\boldsymbol{r},t)$ and $\boldsymbol{M}(\boldsymbol{r},t)$. The pinnacle of these efforts has been the identification of the Poynting vector $\boldsymbol{S}(\boldsymbol{r},t) = \boldsymbol{E}(\boldsymbol{r},t) \times \boldsymbol{H}(\boldsymbol{r},t)$ as the energy flow rate per unit area per unit time, from which all other expressions pertaining to generation, storage, and conversion of electromagnetic energy can be derived in compliance with the law of conservation of energy [9-12]. In contrast, the momentum density of the electromagnetic field within ponderable media has remained unspecified. This indefiniteness has led to ambiguity with regard to radiation forces and electromagnetic momenta [8]. Many theories have sprung up over the past century to interpret the observed phenomena involving radiation pressure in terms of the electromagnetic as well as mechanical momenta of the radiation field; see [13] for an extensive review. (The notation used throughout this paper is the standard notation used in most texts on classical electrodynamics, namely, $\boldsymbol{r}$ and $t$ are the coordinates of an event in space-time; $\boldsymbol{E}$ is the electric field, $\boldsymbol{H}$ the magnetic field, $\boldsymbol{D}$ the electric displacement, $\boldsymbol{B}$ the magnetic induction; $\rho_{\text{free}}$ and $\boldsymbol{J}_{\text{free}}$ are the densities of free charge and free current, respectively; $\boldsymbol{P}$ is the polarization density, $\boldsymbol{M}$ the magnetization density, $\boldsymbol{S}$ the Poynting vector; $c$ is the speed of light in vacuum, $Z_o$ the impedance of free space, $\varepsilon_o$ and $\mu_o$ the permittivity and permeability of free space, and so on.)

In a recent paper [8] we traced the problems associated with the momentum of light in ponderable media to this under-specification on the part of the classical theory. To remedy the shortcoming, we postulated that the electromagnetic momentum density under all circumstances must be given by the Abraham expression $\boldsymbol{p}_{EM} = \boldsymbol{S}(\boldsymbol{r},t)/c^2$, where $\boldsymbol{S}(\boldsymbol{r},t)$ is the Poynting vector and $c$ is the speed of light in vacuum. This expression is thus expected to give the correct electromagnetic momentum density (at point $\boldsymbol{r}$ in space and instant $t$ in time) not only in vacuum, but also within ponderable media, irrespective of whether such media are homogeneous, isotropic, linear, dispersive, absorptive, magnetic, etc. In addition, the aforesaid momentum density does not depend on whether the electromagnetic field is propagating or evanescent, static or dynamic, etc. We have shown under a variety of circumstances – circumstances that have been amenable to analytical and/or numerical investigation – that, in conjunction with Maxwell's macroscopic equations and a generalized form of the Lorentz law of force, the postulated momentum density is consistent with the laws of conservation of linear and angular momentum [14-25]. Specifically, we have demonstrated that, at each and every instant $t$, the time-rate-of-change of the total electromagnetic momentum of a closed system is precisely equal in magnitude and opposite in sign to the total force exerted on the ponderable media within the system.

It remains to explain the curious circumstance that, in certain reported experiments, the measured force of radiation appears to be consistent with the Minkowski momentum [26-28]. In the language of quantum optics, a photon of energy $\hbar\omega$ within a transparent host of refractive index $n_1$ is found to carry a momentum equal to $n_1\hbar\omega/c$, whereas the corresponding Abraham momentum is known to be $\hbar\omega/(n_1 c)$. The goal of the present paper is to show that the appearance of the Minkowski momentum in the reported experiments is *not* indicative of the actual momentum carried by photons within the dielectric; rather, the observed momentum-transfer is a simple consequence of the reduced impedance-mismatch (which happens to be proportional to $n_1$) between the host dielectric of refractive index $n_1$ and the reflective/absorptive medium under investigation. This reduction of impedance-mismatch occurs in such a way as to convey the false impression that, somehow, before and after interacting with the mirror, the photons (within their dielectric host) possess the Minkowski momentum.

In the case of the photon drag effect [28], we presented the corresponding analysis in an earlier paper [15], which will not be repeated here. The concern of the present paper is the force of radiation on an immersed absorber/reflector. The case of illumination by a continuous-wave (cw) light beam will be treated in Sec. 2, followed by an analysis of pulsed illumination in Sec. 3. The entire discussion will be based on the classical electrodynamics of Maxwell and Lorentz. No substantive approximations will be made, lest doubt be cast on the results of mathematical derivations from foundational equations. Aside from the trivial normalization by $\hbar\omega$ of the electromagnetic wave energy (which is done solely for pedagogical reasons), we will not invoke quantum-mechanical arguments, nor borrow mathematical tools from the arsenal of quantum electrodynamics. Standard Fourier transform theory and calculus are all that is needed for the analyses presented in these pages. The results of our Finite Difference Time Domain (FDTD) numerical simulations with emphasis on the effect of focusing a light beam onto a submerged mirror will be described in Sec. 4. Concluding remarks appear in Sec. 5.

**2. Submerged mirror under cw illumination**. Figure 1 shows a linearly-polarized, monochromatic plane-wave propagating along the $z$-axis within a transparent dielectric medium of refractive index $n_1$. At $z=0$ the beam encounters an absorbing medium whose complex refractive index is $n_2+\mathrm{i}\kappa_2$. The interface between the two media is flat and smooth, and incidence is normal. The Fresnel reflection and transmission coefficients at the interface are

$$\rho = E_x^{(r)}/E_x^{(i)} = \frac{n_1-n_2-\mathrm{i}\kappa_2}{n_1+n_2+\mathrm{i}\kappa_2}, \tag{1a}$$

$$\tau = E_x^{(t)}/E_x^{(i)} = 1+\rho = \frac{2n_1}{n_1+n_2+\mathrm{i}\kappa_2}. \tag{1b}$$

The absorber becomes a good mirror when $|\rho|^2 \approx 1$. For metallic mirrors this usually occurs because $\kappa_2$ is quite large. Generally speaking, large values of either $n_2$ or $\kappa_2$ (or both) could result in a large value of $|\rho|^2$. It should be noted, however, that another possible high-reflectivity mirror is one that has a very small $n_2$ (ideally $n_2=0$), in which case the reflectance would be nearly 100% irrespective of the value of $\kappa_2$.

Later in the paper we will consider isotropic, linear, and homogeneous media specified by their permittivity $\varepsilon(\omega)$ as well as permeability $\mu(\omega)$. For such media the (complex) refractive index is $n+\mathrm{i}\kappa = \sqrt{\varepsilon\mu}$, while the (complex) admittance is $\eta = \sqrt{\varepsilon/\mu}$. The Fresnel reflection and transmission coefficients for such media will still be given by Eqs. (1), provided that the admittance $\eta$ is substituted everywhere for the refractive index $n+\mathrm{i}\kappa$.

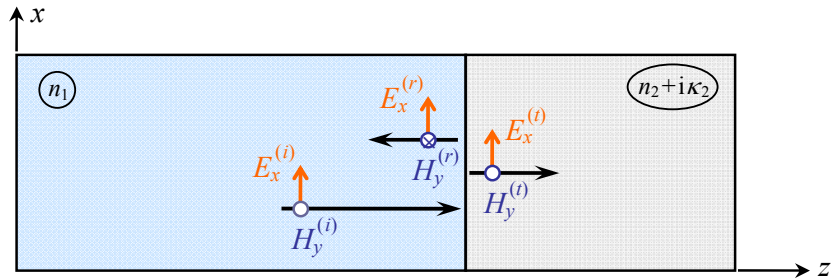

**Figure 1** (color online). A monochromatic plane-wave propagating along the $z$-axis within a homogeneous, isotropic, transparent medium of refractive index $n_1$ is normally incident on a homogeneous, isotropic, absorptive medium of (complex) refractive index $n_2+\mathrm{i}\kappa_2$. The incident, reflected, and transmitted beams are all linearly-polarized along the $x$-axis. The Fresnel reflection and transmission coefficients, $\rho$ and $\tau$, are given by Eqs. (1a) and (1b), respectively.

Denoting the optical frequency by $\omega$, the vacuum wavelength by $\lambda_o$, the speed of light in vacuum by $c=1/\sqrt{\mu_o\varepsilon_o}$, the wave-number by $k_o=\omega/c=2\pi/\lambda_o$, and the impedance of the free-space by $Z_o=\sqrt{\mu_o/\varepsilon_o}$, we may write the incident, reflected, and transmitted electromagnetic fields residing within the two media as follows:

$$\boldsymbol{E}^{(i)}(\boldsymbol{r},t) = E_x^{(i)} \exp[\mathrm{i}k_o(n_1 z - ct)]\hat{\boldsymbol{x}}, \qquad z \leq 0; \quad (2\mathrm{a})$$

$$\boldsymbol{H}^{(i)}(\boldsymbol{r},t) = (n_1/Z_o) E_x^{(i)} \exp[\mathrm{i}k_o(n_1 z - ct)]\hat{\boldsymbol{y}}, \qquad z \leq 0; \quad (2\mathrm{b})$$

$$\boldsymbol{E}^{(r)}(\boldsymbol{r},t) = \rho E_x^{(i)} \exp[\mathrm{i}k_o(n_1 z + ct)]\hat{\boldsymbol{x}}, \qquad z \leq 0; \quad (3\mathrm{a})$$

$$\boldsymbol{H}^{(r)}(\boldsymbol{r},t) = -(n_1/Z_o)\rho E_x^{(i)} \exp[\mathrm{i}k_o(n_1 z + ct)]\hat{\boldsymbol{y}}, \qquad z \leq 0; \quad (3\mathrm{b})$$

$$\boldsymbol{E}^{(t)}(\boldsymbol{r},t) = \tau E_x^{(i)} \exp(-k_o\kappa_2 z)\exp[\mathrm{i}k_o(n_2 z - ct)]\hat{\boldsymbol{x}}, \qquad z \geq 0; \quad (4\mathrm{a})$$

$$\boldsymbol{H}^{(t)}(\boldsymbol{r},t) = [(n_2 + \mathrm{i}\kappa_2)/Z_o]\tau E_x^{(i)} \exp(-k_o\kappa_2 z)\exp[\mathrm{i}k_o(n_2 z - ct)]\hat{\boldsymbol{y}}, \qquad z \geq 0. \quad (4\mathrm{b})$$

The energy flux (per unit area per unit time) in the transparent dielectric is obtained from the time-averaged Poynting vector $\boldsymbol{S}(\boldsymbol{r},t)$, as follows:

$$<\boldsymbol{S}^{(i)}(\boldsymbol{r},t)> = \tfrac{1}{2}\mathrm{Real}\{\boldsymbol{E}^{(i)}(\boldsymbol{r},t) \times \boldsymbol{H}^{*(i)}(\boldsymbol{r},t)\} = \tfrac{1}{2}(n_1/Z_o)|E_x^{(i)}|^2 \hat{\boldsymbol{z}}, \qquad z \leq 0. \quad (5)$$

The time-averaged force per unit area, $<\boldsymbol{F}>$, exerted on the absorbing medium is obtained by integrating the Lorentz force density $(\boldsymbol{P}\cdot\nabla)\boldsymbol{E} + (\partial\boldsymbol{P}/\partial t)\times\mu_o\boldsymbol{H}$ over the penetration depth, namely,

$$<\boldsymbol{F}> = \int_0^\infty <\boldsymbol{F}^{(t)}(\boldsymbol{r},t)>\mathrm{d}z = \int_0^\infty \tfrac{1}{2}\mathrm{Real}\{-\mathrm{i}\omega\varepsilon_o\mu_o(\varepsilon_2-1)\boldsymbol{E}^{(t)}(\boldsymbol{r},t)\times\boldsymbol{H}^{*(t)}(\boldsymbol{r},t)\}\mathrm{d}z$$

$$= \tfrac{1}{2}\mathrm{Real}\{-\mathrm{i}\omega\varepsilon_o\mu_o[(n_2+\mathrm{i}\kappa_2)^2 - 1][(n_2-\mathrm{i}\kappa_2)/Z_o]|\tau E_x^{(i)}|^2 \hat{\boldsymbol{z}}\}\int_0^\infty \exp(-2k_o\kappa_2 z)\mathrm{d}z$$

$$= \frac{\varepsilon_o n_1^2(1+n_2^2+\kappa_2^2)}{(n_1+n_2)^2+\kappa_2^2}|E_x^{(i)}|^2 \hat{\boldsymbol{z}}. \quad (6)$$

It is thus clear that the energy flux of Eq.(5) produces the force on the absorbing medium given by Eq.(6). The energy flux being proportional to the rate of arrival of photons (photon energy = $\hbar\omega$, where $\hbar$ is the reduced Planck constant), we see that the momentum transferred to the mirror/absorber by each incident photon is

$$\frac{<F_z>}{<S_z^{(i)}>/(\hbar\omega)} = \frac{2n_1(1+n_2^2+\kappa_2^2)}{(n_1+n_2)^2+\kappa_2^2}(\hbar\omega/c). \quad (7)$$

A good reflector typically has a large $\kappa_2$ or a large $n_2$ (or both). Thus, so long as the contribution of $n_1$ to the denominator on the right-hand side of Eq.(7) is negligible, the momentum transferred to the absorber by each incident photon is effectively proportional to $n_1$, namely, the (phase) refractive index of the medium hosting the incident photons. As an example, let $n_2+\mathrm{i}\kappa_2=2+\mathrm{i}7$, typical of aluminum at visible wavelengths. The transferred momentum per photon will be $\sim 1.862\hbar\omega/c$ in vacuum ($n_1=1$) and $\sim 2.645\hbar\omega/c$ in a dielectric of refractive index $n_1=1.5$. For the aluminum mirror, therefore, the ratio of the forces is $\sim 1.42$, close to $n_1$. (With a reflector having larger values of $n_2$ and/or $\kappa_2$, this ratio will be even closer to the index of the dielectric host.)

It is clear that, by using a host medium of refractive index $n_1$ rather than the vacuum, one reduces the impedance-mismatch between the host and the mirror, thus increasing the strength of the coupled $E$- and $H$-fields into the mirror, which has the effect of enhancing the momentum transfer per incident photon in proportion to $n_1$. The force on the mirror is not so much a measure of the momentum carried by individual photons within the host as it is a measure of the impedance-mismatch between the host and the mirror. (It is possible, of course, as we have pointed out elsewhere [20], to have a special mirror for which $n_2=0$ and $0<\kappa_2\ll 1$. Under such circumstances, the transferred momentum per photon will be inversely proportional to $n_1$, which is characteristic of the Abraham momentum of the photon within the host medium.)

In another extreme scenario, one may have $n_1=n_2$ while $\kappa_2\approx 0$. Equation (7) then predicts that the momentum transferred to the absorber per incident photon is $\sim \tfrac{1}{2}(n_1+n_1^{-1})(\hbar\omega/c)$. The impedance-mismatch is now essentially removed, and the incident light in its entirety enters the absorber and proceeds to get absorbed, albeit slowly, as $\kappa_2$ is assumed to be small. The transferred momentum in this case is seen to be one-half of the Abraham momentum (in a dispersionless medium of refractive index $n_1$) plus one-half of the Minkowski momentum. This, of course, is the *entire* momentum of each incident photon (as there is no reflection at the interface), comprised of both electromagnetic and mechanical contributions.

The above arguments can be readily extended to magnetic media specified by their permeability $\mu(\omega)$ and permittivity $\varepsilon(\omega)$. Let us assume that the incident medium is transparent, having refractive index $n_1=\sqrt{\varepsilon_1\mu_1}$, where both $\varepsilon_1$ and $\mu_1$ are real-valued, while the absorbing medium has a complex refractive index $n_2=\sqrt{\varepsilon_2\mu_2}$, where either $\varepsilon_2$ or $\mu_2$ or both are complex. Maxwell's equations then reveal that the magnetic fields in Eqs.(2b), (3b) and 4(b), instead of being proportional to the refractive index $n$, will be proportional to the corresponding admittance $\eta=\sqrt{\varepsilon/\mu}$. Similarly, in the expressions of the Fresnel reflection and transmission coefficients, Eqs.(1a) and (1b), admittances replace the refractive indices. Finally, when computing $<F>$ in the presence of the magnetization $M$ within the absorptive medium, one must use the complete Lorentz force density expression [8,22-25], namely,

$$F(r,t) = (P \cdot \nabla)E + (M \cdot \nabla)H + (\partial P/\partial t) \times \mu_\mathrm{o} H - (\partial M/\partial t) \times \varepsilon_\mathrm{o} E. \tag{8}$$

Straightforward calculations then yield

$$\frac{<F_z>}{<S_z^{(i)}>/(\hbar\omega)} = \frac{2\eta_1(1+|\eta_2|^2)}{|\eta_1+\eta_2|^2}(\hbar\omega/c). \tag{9}$$

Once again, the previous arguments apply, namely, that the reduction of the impedance-mismatch between the incidence medium and the absorber/reflector is responsible for the coefficient $\eta_1=\sqrt{\varepsilon_1/\mu_1}$ appearing in the numerator on the right-hand side of Eq.(9). Whenever $\eta_2$ in the denominator dominates $\eta_1$, the transferred momentum per photon will be proportional to $\eta_1$, which has a positive value even in a negative-index medium.

**3. Submerged mirror under pulsed illumination**. Consider a finite-duration, finite-diameter light pulse propagating in a transparent medium of refractive index $n_1$, as shown in Fig.2. The beam diameter is sufficiently large (and its cross-sectional phase and amplitude profiles sufficiently uniform) that one may safely ignore the diffraction effects. The total energy content of the pulse will be denoted by $\mathscr{E}_\mathrm{pulse}$. The *E*- and *H*-fields within the homogeneous and isotropic host media will be expressed in terms of a plane-wave spectrum of spatio-temporal frequencies $(k_x, k_y, \omega)$, with the spatial-frequency content of the pulse confined to a small region of the *k*-space in the vicinity of $(k_x, k_y) = (0, 0)$.

$$E(r,t) = \tfrac{1}{2}\iiint \mathscr{E}(k_x, k_y, \omega)\exp[\mathrm{i}(k_x x + k_y y + k_z z - \omega t)]\mathrm{d}k_x \mathrm{d}k_y \mathrm{d}\omega, \tag{10a}$$

$$H(r,t) = \tfrac{1}{2}\iiint \mathscr{H}(k_x, k_y, \omega)\exp[\mathrm{i}(k_x x + k_y y + k_z z - \omega t)]\mathrm{d}k_x \mathrm{d}k_y \mathrm{d}\omega. \tag{10b}$$

In non-magnetic media, where the relative permittivity $\mu(\omega)$ is unity, we have

$$k_z = (\omega/c)\sqrt{\varepsilon(\omega) - (ck_x/\omega)^2 - (ck_y/\omega)^2}, \tag{11}$$

where $\varepsilon(\omega)$ is the relative permittivity of the medium, with the proviso that $\varepsilon(-\omega) = \varepsilon^*(\omega)$. Assuming dispersionless behavior, $\varepsilon_1(\omega) = n_1^2$ in the case of the dielectric host, while $\varepsilon_2(\omega) = (n_2 + \mathrm{i}\kappa_2)^2$ in the case of the absorbing medium ($\kappa_2 > 0$ when $\omega > 0$). For the fields to be real-valued it is necessary and sufficient that their Fourier transforms be Hermitian, that is,

$$\mathscr{E}(k_x, k_y, \omega) = \mathscr{E}^*(-k_x, -k_y, -\omega), \qquad \mathscr{H}(k_x, k_y, \omega) = \mathscr{H}^*(-k_x, -k_y, -\omega). \tag{12}$$

Moreover, if the beam's cross-section in the *xy*-plane is required to be symmetric with respect to the origin, that is, if the field amplitudes are to remain intact upon switching $(x, y)$ to $(-x, -y)$, we must have

$$\mathscr{E}(k_x, k_y, \omega) = \mathscr{E}(-k_x, -k_y, \omega), \qquad \mathscr{H}(k_x, k_y, \omega) = \mathscr{H}(-k_x, -k_y, \omega). \tag{13}$$

The $\mathscr{E}$- and $\mathscr{H}$-field amplitudes and the *k*-vector $\boldsymbol{k} = k_x\hat{\boldsymbol{x}} + k_y\hat{\boldsymbol{y}} + k_z\hat{\boldsymbol{z}}$ are related through the Maxwell equations $\nabla \cdot \boldsymbol{E} = 0$ and $\nabla \times \boldsymbol{E} = -\partial\boldsymbol{B}/\partial t$, as follows:

$$\boldsymbol{k} \cdot \boldsymbol{\mathscr{E}}(k_x, k_y, \omega) = k_x\mathscr{E}_x + k_y\mathscr{E}_y + k_z\mathscr{E}_z = 0, \tag{14a}$$

$$\boldsymbol{k} \times \boldsymbol{\mathscr{E}}(k_x, k_y, \omega) = \omega\mu_\mathrm{o}\boldsymbol{\mathscr{H}}(k_x, k_y, \omega). \tag{14b}$$

With the help of Eq. (14b) and the vector identity $\boldsymbol{a} \times (\boldsymbol{b} \times \boldsymbol{c}) = (\boldsymbol{a} \cdot \boldsymbol{c})\boldsymbol{b} - (\boldsymbol{a} \cdot \boldsymbol{b})\boldsymbol{c}$, we now write

$$\boldsymbol{\mathcal{E}}(k_x, k_y, \omega) \times \boldsymbol{\mathcal{H}}^*(k_x, k_y, \omega) = (\omega \mu_0)^{-1} \boldsymbol{\mathcal{E}}(k_x, k_y, \omega) \times [\boldsymbol{k}^* \times \boldsymbol{\mathcal{E}}^*(k_x, k_y, \omega)]$$

$$= (\omega \mu_0)^{-1} \{ |\boldsymbol{\mathcal{E}}(k_x, k_y, \omega)|^2 \boldsymbol{k}^* - [\boldsymbol{k}^* \cdot \boldsymbol{\mathcal{E}}(k_x, k_y, \omega)] \boldsymbol{\mathcal{E}}^*(k_x, k_y, \omega) \}. \quad (15)$$

For the beam thus defined, the incident Poynting vector in the transparent host medium may be written

$$\boldsymbol{S}^{(i)}(\boldsymbol{r}, t) = \boldsymbol{E}^{(i)}(\boldsymbol{r}, t) \times \boldsymbol{H}^{(i)}(\boldsymbol{r}, t) = \tfrac{1}{4} \iiiiii \boldsymbol{\mathcal{E}}^{(i)}(k_x, k_y, \omega) \times \boldsymbol{\mathcal{H}}^{(i)}(k'_x, k'_y, \omega') \exp[\mathrm{i}(k_x + k'_x)x] \exp[\mathrm{i}(k_y + k'_y)y]$$

$$\times \exp[\mathrm{i}(k_z + k'_z)z] \exp[-\mathrm{i}(\omega + \omega')t] \, \mathrm{d}k_x \mathrm{d}k_y \mathrm{d}\omega \mathrm{d}k'_x \mathrm{d}k'_y \mathrm{d}\omega'. \quad (16)$$

Integrating $S_z(\boldsymbol{r}, t)$ over the beam's cross-sectional area in the $xy$-plane and over all time, then using the identity

$$\int_{-\infty}^{\infty} \exp[\mathrm{i}(k + k')\zeta] \, \mathrm{d}\zeta = 2\pi \delta(k + k'), \quad (17)$$

where $\delta(k)$ is Dirac's delta function, the pulse's total energy content turns out to be

$$\mathcal{E}_{\text{pulse}} = \iiint_{-\infty}^{\infty} S_z^{(i)}(x, y, z = z_0, t) \, \mathrm{d}x \mathrm{d}y \mathrm{d}t = \tfrac{1}{4} \iiint_{-\infty}^{\infty} [\boldsymbol{\mathcal{E}}^{(i)}(k_x, k_y, \omega) \times \boldsymbol{\mathcal{H}}^{(i)*}(k_x, k_y, \omega)]_z \, \mathrm{d}k_x \mathrm{d}k_y \mathrm{d}\omega. \quad (18)$$

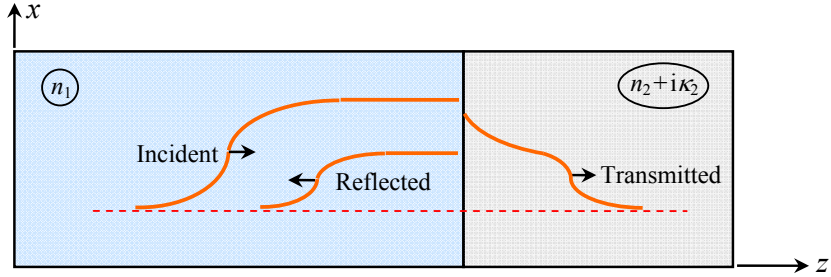

**Figure 2** (color online). A finite-duration, finite-diameter light pulse propagating along the $z$-axis within a homogeneous, isotropic, transparent medium of refractive index $n_1$ is normally incident on a homogeneous, isotropic, absorptive medium of complex refractive index $n_2 + \mathrm{i}\kappa_2$. The Fresnel reflection coefficient at the interface is $\rho$, and the energy content of the incident pulse is $\mathcal{E}_{\text{pulse}}$. While the reflected pulse continues to carry its total energy of $|\rho|^2 \mathcal{E}_{\text{pulse}}$, the transmitted pulse is attenuated as it penetrates the absorptive medium. Eventually, the entire transmitted pulse is absorbed by the medium, transferring its electromagnetic momentum to its temporary host. This, however, is *not* the only momentum transferred to the absorptive medium; the Lorentz force of the light exerted on this medium before the fields vanish also contributes to the transferred momentum.

We now substitute in Eq. (18) from Eq. (15), where we use the fact that $(k_x, k_y)$ is in the vicinity of $(0, 0)$, so that, in accordance with Eq. (14a), the $\mathcal{E}_z$ component of the field can be ignored; moreover, from Eq. (11) we have $k_z \approx n_1 \omega/c$. We find

$$\mathcal{E}_{\text{pulse}} \approx [n_1/(4Z_0)] \iiint_{-\infty}^{\infty} |\boldsymbol{\mathcal{E}}^{(i)}(k_x, k_y, \omega)|^2 \, \mathrm{d}k_x \mathrm{d}k_y \mathrm{d}\omega. \quad (19)$$

Next we calculate the total force $\boldsymbol{F}$ of the light pulse on the semi-infinite absorptive medium by integrating the force density, $(\boldsymbol{P} \cdot \boldsymbol{\nabla})\boldsymbol{E} + (\partial \boldsymbol{P}/\partial t) \times \mu_0 \boldsymbol{H}$, over the spatial coordinates $x, y$ and $z$. Since $(k_x, k_y)$ is in the vicinity of $(0, 0)$, and also because the beam's cross-section is assumed to be symmetric in the $xy$-plane with respect to the origin, the contribution of $(\boldsymbol{P} \cdot \boldsymbol{\nabla})\boldsymbol{E}$, integrated over the beam's cross-section, will be negligible. We may also suppose that all the plane-waves enter the absorptive medium essentially at normal incidence and, therefore, with the same transmission coefficient $\tau$ as in Eq. (1b). The density of the force experienced by the absorptive medium is

$$\boldsymbol{F}^{(t)}(\boldsymbol{r}, t) = \tfrac{1}{4} \iiiiii_{-\infty}^{\infty} -\mathrm{i} \omega \mu_0 \varepsilon_0 [\varepsilon_2(\omega) - 1] \boldsymbol{\mathcal{E}}^{(t)}(k_x, k_y, \omega) \times \boldsymbol{\mathcal{H}}^{(t)}(k'_x, k'_y, \omega') \exp[\mathrm{i}(k_x + k'_x)x] \exp[\mathrm{i}(k_y + k'_y)y]$$

$$\times \exp[\mathrm{i}(k_z + k'_z)z] \exp[-\mathrm{i}(\omega + \omega')t] \, \mathrm{d}k_x \mathrm{d}k_y \mathrm{d}\omega \mathrm{d}k'_x \mathrm{d}k'_y \mathrm{d}\omega'. \quad (20)$$

The above force density, when integrated over space and time will, with the help of Eqs. (15) and (17), yield

$$\iiint_{-\infty}^{\infty}\int_{z=0}^{\infty} \boldsymbol{F}^{(t)}(\boldsymbol{r},t)\,\mathrm{d}t\,\mathrm{d}x\,\mathrm{d}y\,\mathrm{d}z$$

$$= \tfrac{1}{4}\iiint_{-\infty}^{\infty} -\mathrm{i}\omega\mu_0\varepsilon_0[\varepsilon_2(\omega)-1]\left\{\int_{z=0}^{\infty}\exp[\mathrm{i}(k_z-k_z^*)z]\mathrm{d}z\right\}\boldsymbol{\mathcal{E}}^{(t)}(k_x,k_y,\omega)\times\boldsymbol{\mathcal{H}}^{(t)*}(k_x,k_y,\omega)\,\mathrm{d}k_x\mathrm{d}k_y\mathrm{d}\omega$$

$$= \tfrac{1}{4}\varepsilon_0\iiint_{-\infty}^{\infty}\{[\varepsilon_2(\omega)-1]/(k_z-k_z^*)\}\{|\boldsymbol{\mathcal{E}}^{(t)}(k_x,k_y,\omega)|^2\boldsymbol{k}^* - [\boldsymbol{k}^*\cdot\boldsymbol{\mathcal{E}}^{(t)}(k_x,k_y,\omega)]\boldsymbol{\mathcal{E}}^{(t)*}(k_x,k_y,\omega)\}\mathrm{d}k_x\mathrm{d}k_y\mathrm{d}\omega. \qquad (21)$$

As before, since $k_x, k_y$ are close to zero, we ignore the $\mathcal{E}_z$ component of the field and proceed to approximate $k_z$ with $(\omega/c)\sqrt{\varepsilon_2(\omega)}$. The time-integrated force along the $z$-axis experienced by the absorptive medium becomes

$$\iiint_{-\infty}^{\infty}\int_{z=0}^{\infty} F_z^{(t)}(\boldsymbol{r},t)\,\mathrm{d}t\,\mathrm{d}x\,\mathrm{d}y\,\mathrm{d}z \approx \tfrac{1}{4}\varepsilon_0|\tau|^2 \iiint_{-\infty}^{\infty} \frac{[\varepsilon_2(\omega)-1]\sqrt{\varepsilon_2^*(\omega)}}{\sqrt{\varepsilon_2(\omega)}-\sqrt{\varepsilon_2^*(\omega)}}|\boldsymbol{\mathcal{E}}^{(i)}(k_x,k_y,\omega)|^2\,\mathrm{d}k_x\mathrm{d}k_y\mathrm{d}\omega. \qquad (22)$$

In the above integrand, the coefficient of $|\boldsymbol{\mathcal{E}}^{(i)}|^2$ becomes its own conjugate when $\omega$ changes sign to $-\omega$. Since the $\mathcal{E}$-field intensity remains the same under this sign change, one is justified in replacing the coefficient with its real-part, namely, $\tfrac{1}{2}[1+|\varepsilon_2(\omega)|]$. In the absence of dispersion, this coefficient is independent of $\omega$ and may be taken outside the integral. We thus have

$$\iiint_{-\infty}^{\infty}\int_{z=0}^{\infty} F_z^{(t)}(\boldsymbol{r},t)\,\mathrm{d}t\,\mathrm{d}x\,\mathrm{d}y\,\mathrm{d}z \approx \frac{\varepsilon_0 n_1^2(1+n_2^2+\kappa_2^2)}{2[(n_1+n_2)^2+\kappa_2^2]}\iiint_{-\infty}^{\infty}|\boldsymbol{\mathcal{E}}^{(i)}(k_x,k_y,\omega)|^2\,\mathrm{d}k_x\mathrm{d}k_y\mathrm{d}\omega. \qquad (23)$$

Normalizing the time-integrated force in Eq. (23) by the pulse energy $\mathcal{E}_{\text{pulse}}$ given in Eq. (19) yields the same result as obtained earlier in Eq. (7) for the case of cw illumination. Once again, the momentum per photon transferred by the light pulse to a good reflector is seen to be proportional to $n_1$, not because of an inherent Minkowski momentum of the photons inside the transparent host, but because of a reduced impedance-mismatch between the dielectric host and the absorptive medium.

The above results pertaining to pulsed illumination can be further extended to cover magnetic media specified by their $\varepsilon(\omega)$ and $\mu(\omega)$. The final result turns out to be identical to Eq. (9), which was obtained in Sec. 2 for cw illumination of a magnetic absorber/reflector through a transparent magnetic dielectric. Equation (9) is thus seen to be quite general, applicable under pulsed as well as cw illumination to magnetic and non-magnetic media alike.

It must be pointed out that the treatment of the present section is limited in scope to fairly long pulses traveling in homogeneous, isotropic, linear, non-dispersive media. The case of very short pulses propagating in inhomogeneous, anisotropic, dispersive, and absorptive media is much more complicated and requires a treatment such as that given by Cs. Ferencz et al [29].

The momentum of a light pulse traveling within a transparent host consists of both electromagnetic and mechanical components, the total momentum per photon being $\tfrac{1}{2}(\eta_1+\eta_1^{-1})\hbar\omega/c$ [21,22]. The force experienced by a good reflector submerged in a transparent medium, however, is generally greater than that indicated by the sum of the incident and reflected momenta, for the following reason: Upon reflection, the incident and reflected pulses overlap for some period of time. In the region of overlap, there will be interference, and a net force is exerted on the transparent host (the incidence medium) whose direction is typically opposite the direction of the force exerted on the mirror [21]. Consequently, the net force experienced by the mirror in accordance with Eq. (9) will be greater than that expected from a simple argument based solely on the conservation of the pulse's total (i.e., electromagnetic plus mechanical) momentum. In the Appendix we describe a method of calculating the mechanical momentum transferred to the host medium as a result of the overlap between incident and reflected pulses in the dielectric region immediately preceding the reflector.

**4. Computer simulations**. Finite Difference Time Domain (FDTD) simulations are a powerful tool for computing numerically exact solutions to Maxwell's equations. With reference to Fig. 3, we simulated the case of a linearly polarized Gaussian beam of vacuum wavelength $\lambda_0 = 650$ nm, focused via an aberration-free lens into a liquid droplet placed atop the flat surface of an absorber/reflector. A cylindrical lens will produce a line focus along the $y$-axis, in which case the computations can be carried out in two spatial dimensions (2D). A spherical lens will produce a 3D focused spot having a nearly circular cross-sectional profile. In both cases the radiation pressure on the absorber/reflector may be calculated as a function of the full-width-at-half-maximum-amplitude (FWHM) of the focused Gaussian spot.

For a focused spot having FWHM greater than about 2 μm, the force of radiation is found to be in agreement with the plane-wave result predicted by Eq. (7). As the spot diameter shrinks, however, we find that the radiation pressure declines rapidly. Figure 4 shows computed plots of normalized force versus the spot's FWHM diameter for a light beam focused onto a silver mirror through a liquid droplet of refractive index $n_1=1.5$. Both cases of 2D and 3D focusing are depicted in the figure, with 3D focusing seen to result in a more rapid decline in radiation pressure as the spot size shrinks below about 1 μm. For 2D focusing, the computed radiation pressure on the mirror is shown in Fig. 4 for both s- and p-polarized light, i.e., with the $E$-field being, respectively, parallel and perpendicular to the line-focus. These plots reveal that, with a decreasing spot diameter, the normalized force of a p-polarized beam – focused onto the silver mirror through a cylindrical lens – drops faster than that of an s-polarized beam. This is understandable considering that (i) a sharply-focused light spot consists of a large number of plane-waves arriving at the mirror surface at oblique incidence, and (ii) the Fresnel reflection coefficient of the mirror at oblique incidence is greater for s-light than for p-light.

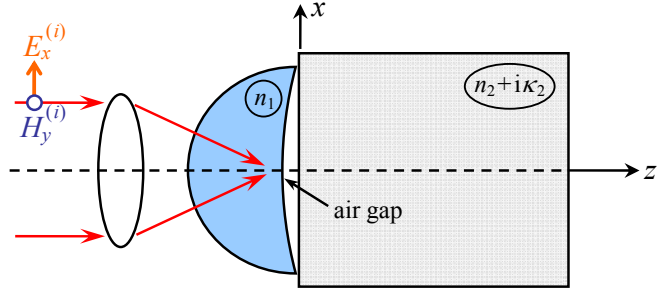

**Figure 3** (color online). A droplet of liquid having refractive index $n_1=1.5$ is placed atop an absorber/reflector having a complex refractive index $n_2+\mathrm{i}\kappa_2=0.0608+\mathrm{i}4.6194$ (dielectric constant $\varepsilon=-21.3353+\mathrm{i}0.56186$). A Gaussian beam of wavelength $\lambda_\mathrm{o}=650$ nm, linearly-polarized along the $x$-axis, is brought to focus at a distance of 100 nm before the solid-liquid interface at $z=0$. In our computer simulations the focusing power of the lens was adjusted to allow the full-width-at-half-maximum amplitude of the focused (Gaussian) spot to vary in the range from 0.3 μm to 8 μm. With a cylindrical lens, the focused beam will be invariant along the $y$-axis (2D simulations), whereas a spherical lens produces a focused spot with a nearly circular cross-section (3D simulations). A 2 nm air gap is introduced at the solid-liquid interface to facilitate the computation of the Lorentz force exerted on the bound charges at the absorber/reflector surface.

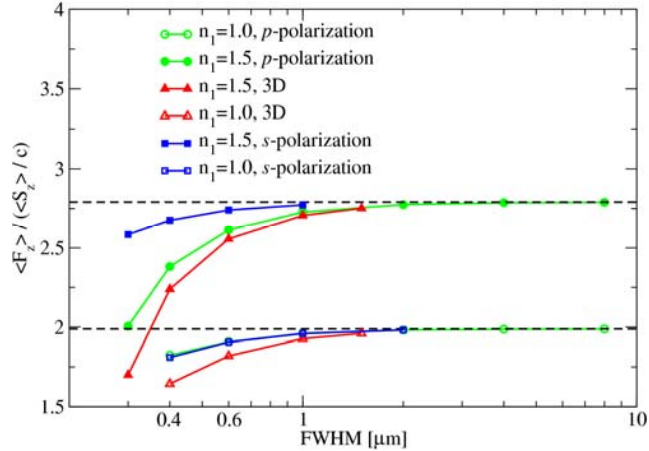

**Figure 4** (color online). Plots of $<F_z>/(<S_z>/c)$ versus the focused spot's FWHM in the system depicted in Fig. 3. The solid squares (blue) and solid circles (green) represent the case of focusing through a cylindrical lens (2D), whereas the solid triangles (red) correspond to focusing through a spherical lens (3D). In both cases, in the limit of large beam-diameter (i.e., FWHM > 2.0 μm), the normalized force approaches the value of 2.82 predicted by Eq. (7). Increasing the focusing power of the lens causes the spot-size to shrink, thus resulting in a reduced force (per photon) on the submerged mirror. For a given focusing power, the spherical lens is more effective at reducing the radiation pressure than the cylindrical lens. In the absence of the liquid droplet (i.e., $n_1=1.0$), when the light is focused directly onto the mirror surface, the normalized force drops below the plane-wave limit of 1.989 [predicted by Eq. (7)] when the FWHM spot size becomes less than $\sim 2\lambda_\mathrm{o}$. The open squares (blue) and open circles (black) now represent 2D focusing, while the open triangles (red) correspond to 3D focusing. The decline of radiation pressure with a shrinking spot size in the absence of a droplet is not quite as dramatic as it is when the light is focused through the liquid.

In our FDTD simulations, we introduced a 2 nm air gap between the liquid droplet and the mirror surface in order to facilitate the computation of the Lorentz force on the bound surface charges of the mirror; see Fig. 3. Since the gap is much smaller than a wavelength ($\lambda_o = 650$ nm), its effect on reflected and transmitted $E$- and $H$-fields should be negligible. For a finite-diameter incident beam, surface charges (or electric dipoles) appear on both sides of the solid-liquid interface. The artificial device of an air gap is thus helpful in separating these charges from each other in order to keep track of the Lorentz force on either side of the interface without introducing an unacceptable error in the final numerical results.

In the case of a plane-wave at normal incidence, surface charges are absent and the seat of the Lorentz force is solely within the skin depth of the mirror. For a normally-incident plane-wave, we computed the normalized force on the silver mirror both with and without the 2 nm air gap; the presence of the gap resulted in a change of less than 1% in the computed force. Moreover, when the step size of our FDTD mesh was reduced below 2 nm, the air gap could be reduced as well (since the air gap must be at least equal to one mesh step); in such cases the difference in the computed values of the force with and without the gap was seen to decrease in proportion to the mesh step to values below 1%.

The reduction of radiation pressure with a decreasing beam diameter may be attributed to a reduced incident momentum (per photon) as well as a weaker coupling of the electromagnetic field into the mirror. This effect is not limited to submerged mirrors and may be observed in the case of air-incident focused beams as well. The lower plots in Fig. 4 correspond to air-incident focused beams produced by a cylindrical lens (green open circles, blue open squares) and a spherical lens (red open triangles). Once again, it is seen that for a large-diameter beam (i.e., FWHM $> 2\,\mu$m) the computed normalized force agrees with the result predicted by Eq.(7) when $n_1 = 1.0$, namely, $\langle F_z \rangle / (\langle S_z \rangle / c) = 1.989$. The radiation pressure declines with a decreasing spot size, albeit not quite as fast as it does in the case of a submerged mirror. This reduction is primarily due to the fact that, in vacuum, the electromagnetic momentum (per photon) of a focused spot is less than that of a plane-wave. When a plane-wave is focused (in vacuum) onto a perfect mirror, upon reflection from the mirror and a second passage through the focusing lens (this time in the reverse direction, of course) the entire momentum of the plane-wave will have been reversed. Under these circumstances, the fraction of the plane-wave's electromagnetic momentum that is *not* acquired by the mirror will, in fact, be transferred to the focusing/collimating lens.

**5. Concluding remarks**. In this paper we examined the radiation pressure experienced by an absorber/reflector immersed in a transparent, homogeneous, isotropic medium. The pressure was found to depend on the admittance $\eta_2(\omega) = \sqrt{\varepsilon_2(\omega)/\mu_2(\omega)}$ of the absorber/reflector as well as on $\eta_1(\omega) = \sqrt{\varepsilon_1(\omega)/\mu_1(\omega)}$ of the transparent host, irrespective of whether the incident radiation is pulsed or cw. [In the special case of a non-magnetic host, where $\mu_1(\omega) = 1$, the admittance of the host reduces to its phase refractive index $n_1(\omega) = \sqrt{\varepsilon_1(\omega)}$.] The general expression for the transferred momentum per incident photon of frequency $\omega$ is given by Eq.(9), where the incident beam is assumed to be quasi-monochromatic and collimated (with a diameter that is substantially greater than a wavelength), arriving at normal incidence at the flat interface between the transparent host and the absorbing/reflecting medium. When the relative values of $\eta_1$ and $\eta_2$ are such that, on the right-hand side of Eq.(9), the contribution of $\eta_1$ to the denominator is negligible (e.g., when $\eta_2$ has a large imaginary part, as in good metallic reflectors), the transferred momentum per incident photon is seen to become simply proportional to $\eta_1$ (or, for a non-magnetic host, to $n_1$). Under such circumstances, one is tempted to say that the photons inside the transparent host carry the Minkowski momentum, although the proper interpretation is that a reduced impedance-mismatch has enhanced the coupling of the incident photons to the absorber/reflector.

In the 1953 experiments of Jones and Richards on submerged mirrors [26], followed by the refined 1977 experiments of Jones and Leslie [27], the mirror was highly reflective, and the radiation pressure was found to be proportional to the phase refractive index of the host liquid. These results were interpreted as providing strong evidence in support of a Minkowski momentum for photons traveling within transparent dielectrics. Our analysis in the preceding sections, however, suggests a reinterpretation in terms of a reduced impedance-mismatch.

A fruitful extension of the experimental work in this area is suggested by Eq.(9), where it seems desirable to examine situations in which $\eta_1$ makes a substantial contribution to the denominator. Another possible avenue for further investigation would be the conduction of experiments involving radiation pressure on submerged absorbers/reflectors where the incident beam is highly focused (or highly divergent), thereby necessitating the modification of assumptions that have led to Eq.(9). The computer simulation results presented in Sec. 4 predict a substantial reduction in the radiation pressure per photon as the diameter of the focused light spot drops below about one wavelength.

## Appendix

When a pulse of light traveling within a transparent dielectric medium of refractive index $n_1$ is reflected from a (submerged) mirror, the overlap between the incident and reflected pulses within the dielectric host sets up interference fringes which exert a Lorentz force on the dielectric medium. Here we calculate the net momentum transferred to the host medium in consequence of this overlap and the resulting interference between the two waves.

Let the incident beam's cross-section in the $xy$-plane be sufficiently large and uniform, so that the dependence of the fields on the $x$ and $y$ coordinates may be ignored. Assuming that the mirror is located at $z=0$, while, at $t=0$ the pulse is centered at $z=z_0$ (here $z_0$ is large and negative), the incident pulse's $E$ and $H$ fields may be written

$$E_x^{(i)}(z,t) = \tfrac{1}{2} \int_{-\infty}^{\infty} \mathcal{E}_x(\omega) \exp[\mathrm{i}(n_1\omega/c)(z-z_0) - \mathrm{i}\omega t] \mathrm{d}\omega, \tag{A1a}$$

$$H_y^{(i)}(z,t) = \tfrac{1}{2} \int_{-\infty}^{\infty} (n_1/Z_0) \mathcal{E}_x(\omega) \exp[\mathrm{i}(n_1\omega/c)(z-z_0) - \mathrm{i}\omega t] \mathrm{d}\omega. \tag{A1b}$$

To simplify the analysis, we assume that, within the light pulse's spectral bandwidth, i.e., $|\omega-\omega_0|<\Delta\omega$, $n_1$ is frequency-independent. The dielectric medium is thus free from dispersion, and its real-valued $n_1$, representing both phase and group refractive indices, is greater than unity. Following the procedure that led to Eq.(19), the energy content (per unit cross-sectional area) of the pulse is found to be

$$\mathcal{E}_{\mathrm{pulse}} = [n_1/(4Z_0)] \int_{-\infty}^{\infty} |\mathcal{E}_x(\omega)|^2 \mathrm{d}\omega. \tag{A2}$$

Next we assume the pulse is reflected from a flat mirror whose Fresnel reflection coefficient (within the host dielectric) is $\rho(\omega) = |\rho(\omega)| \exp[\mathrm{i}\phi(\omega)]$. The reflected pulse's $E$ and $H$ fields are given by

$$E_x^{(r)}(z,t) = \tfrac{1}{2} \int_{-\infty}^{\infty} \rho(\omega) \mathcal{E}_x(\omega) \exp[\mathrm{i}(n_1\omega/c)(-z-z_0) - \mathrm{i}\omega t] \mathrm{d}\omega, \tag{A3a}$$

$$H_y^{(r)}(z,t) = -\tfrac{1}{2}(n_1/Z_0) \int_{-\infty}^{\infty} \rho(\omega) \mathcal{E}_x(\omega) \exp[\mathrm{i}(n_1\omega/c)(-z-z_0) - \mathrm{i}\omega t] \mathrm{d}\omega. \tag{A3b}$$

The Lorentz force density $\boldsymbol{F}(\boldsymbol{r},t) = (\partial \boldsymbol{P}/\partial t) \times \mu_0 \boldsymbol{H}$ of the superposed incident and reflected pulses is thus written

$$F_z(z,t) = \left\{\tfrac{\partial}{\partial t} \varepsilon_0 (n_1^2-1)[E_x^{(i)}(z,t) + E_x^{(r)}(z,t)]\right\} [\mu_0 H_y^{(i)}(z,t) + \mu_0 H_y^{(r)}(z,t)]$$

$$= \tfrac{1}{4}\varepsilon_0(n_1^2-1)(\mu_0 n_1/Z_0) \iint_{-\infty}^{\infty} (-\mathrm{i}\omega) \mathcal{E}_x(\omega) \mathcal{E}_x(\omega') \exp[-\mathrm{i}(n_1/c)(\omega+\omega')z_0] \exp[-\mathrm{i}(\omega+\omega')t]$$

$$\times \{\exp[\mathrm{i}(n_1\omega/c)z] + \rho(\omega)\exp[-\mathrm{i}(n_1\omega/c)z]\} \{\exp[\mathrm{i}(n_1\omega'/c)z] - \rho(\omega')\exp[-\mathrm{i}(n_1\omega'/c)z]\} \mathrm{d}\omega \mathrm{d}\omega'. \tag{A4}$$

The net momentum transferred to any location on the $z$-axis is found by integrating $F_z(z,t)$ of Eq.(A4) over time. Using the identity $\int_{-\infty}^{\infty} \exp[-\mathrm{i}(\omega+\omega')t] \mathrm{d}t = 2\pi\delta(\omega+\omega')$, followed by carrying out the integration over $\omega'$ yields

$$\int_{-\infty}^{\infty} F_z(z,t) \mathrm{d}t = \tfrac{1}{4}\varepsilon_0(n_1^2-1)(n_1/c) \int_{-\infty}^{\infty} (-\mathrm{i}\omega)|\mathcal{E}_x(\omega)|^2 \{1-|\rho(\omega)|^2 - 2\mathrm{i}|\rho(\omega)|\sin[2(n_1\omega/c)z - \phi(\omega)]\} \mathrm{d}\omega$$

$$= -\tfrac{1}{2}\varepsilon_0(n_1^2-1)(n_1/c) \int_{-\infty}^{\infty} \omega|\mathcal{E}_x(\omega)|^2 |\rho(\omega)| \sin[2(n_1\omega/c)z - \phi(\omega)] \mathrm{d}\omega. \tag{A5}$$

In the above equation, the first two terms within the curly brackets make no contribution to the integral because $\omega|\mathcal{E}_x(\omega)|^2 [1-|\rho(\omega)|^2]$ is an odd function of $\omega$. Thus the incident and reflected pulses by themselves do not contribute to the time-integrated force (i.e., mechanical momentum) at any point $z$, the reason being that, at each such point, the momentum given to the host molecules by the leading edge of each pulse is taken away by the trailing edge of the same pulse. Within the confines of each pulse, of course, individual molecules do carry a mechanical momentum; what Eq.(A5) says, however, is that, in the fullness of time, the net momentum transferred by individual pulses to each molecule amounts to zero. The sole contribution to the net momentum, according to Eq.(A5), arises from the cross-talk between the incident and reflected pulses, which occur, of course, when these pulses overlap.

Note in Eq. (A5) that, for points located near the mirror (i.e., $z$ close to zero), the momentum transfer tends to be significant. As one moves away from the mirror, $z$ becomes large (and negative), causing the sinusoid to oscillate more and more rapidly with changes of $\omega$. For sufficiently large $z$, i.e., outside the overlap region, these oscillations cancel each other out and the integral vanishes. To find the total mechanical momentum transferred to the host medium in the process of reflection from the submerged mirorr (which is located at $z=0$), we integrate Eq. (A5) over $z$ from $-\infty$ to 0. Due to the aforementioned oscillations of the sinusoid at large $z$, the integrated function goes to zero at $z=-\infty$. The total integrated momentum is thus found to be

$$\int_{-\infty}^{0} dz \int_{-\infty}^{\infty} F_z(z,t) dt = \tfrac{1}{4}\varepsilon_0 (n_1^2 - 1) \int_{-\infty}^{\infty} |\mathcal{E}_x(\omega)|^2 |\rho(\omega)| \cos\phi(\omega) d\omega. \quad (A6)$$

Assuming the reflection coefficient $\rho(\omega)$ is fairly uniform across the spectral bandwidth (centered at $\omega = \omega_o$), we bring $|\rho(\omega)|\cos\phi(\omega)$ outside the integral, then use Eq. (A2) to express the final result in terms of the incident pulse energy. We obtain

$$\textit{Total momentum transferred to dielectric host} \approx |\rho(\omega_o)|\cos\phi(\omega_o)(n_1 - n_1^{-1})\mathcal{E}_{\text{pulse}}/c. \quad (A7)$$

Clearly, the momentum transferred to the dielectric host is proportional to the difference between the Minkowski and Abraham momenta of the incident pulse, $n_1 \mathcal{E}_{\text{pulse}}/c$ and $\mathcal{E}_{\text{pulse}}/(n_1 c)$, respectively. For a perfect reflector, of course, $|\rho(\omega_o)| = 1$. If $\phi(\omega_o) \approx \pi$, which is the case for a good metallic reflector, then the direction of the momentum transferred to the transparent host is away from the mirror. Considering that the incident and reflected pulses each carry a total momentum of $\tfrac{1}{2}(n_1 + n_1^{-1})\mathcal{E}_{\text{pulse}}/c$, the Minkowski momentum of $2n_1\mathcal{E}_{\text{pulse}}/c$ transferred to the mirror is seen to be consistent with the principle of conservation of momentum.

In contrast to the conventional mirror discussed in the present paper, one could construct a nearly perfect reflector whose $\phi(\omega_o) \approx 0$. (This could be achieved, for example, with a dielectric stack of alternating high- and low-refractive-index layers, where some of the layer thicknesses are judiciously chosen to deviate from the standard value of one-quarter-of-one-wavelength.) Equation (A7) now indicates that the transferred momentum to the dielectric host would point *toward* the mirror. A simple balance-of-momentum argument reveals that, under such circumstances, the momentum transferred to the (unconventional) mirror will be $2\mathcal{E}_{\text{pulse}}/(n_1 c)$, as though the photons within the dielectric host carried the Abraham momentum. This prediction may also be confirmed by a direct calculation of the Lorentz force experienced by the (unconventional) mirror [20]. The photons, of course, always carry the *same* momentum, irrespective of the nature of the mirror used in the experiment. What is different in each case is the magnitude and direction of the momentum taken up by the dielectric host (within which the mirror is submerged). This last point is made abundantly clear by the dependence of the transferred momentum in Eq. (A7) on the Fresnel reflection coefficient $\rho(\omega_o)$ of the mirror.

**Acknowledgements.** This work has been supported by the Air Force Office of Scientific Research (AFOSR) under contract number FA 9550-04-1-0213.

# References

1. A. Ashkin and J. Dziedzic, "Radiation pressure on a free liquid surface," Phys. Rev. Lett. **30**, 139-142 (1973).
2. J. P. Gordon, "Radiation forces and momenta in dielectric media," Phys. Rev. A **8**, 14-21 (1973).
3. R. Loudon, "Theory of the radiation pressure on dielectric surfaces," J. Mod. Opt. **49**, 821-838 (2002).
4. R. Loudon, "Theory of the forces exerted by Laguerre-Gaussian light beams on dielectrics," Phys. Rev. A **68**, 013806 (2003).
5. M. Padgett, S. Barnett, and R. Loudon, "The angular momentum of light inside a dielectric," J. Mod. Opt. **50**, 1555-1562 (2003).
6. R. Loudon, "Radiation pressure and momentum in dielectrics," Fortschr. Phys. **52**, 1134-1140 (2004).
7. S. M. Barnett and R. Loudon, "On the electromagnetic force on a dielectric medium," *J. Phys. B: At. Mol. Opt. Phys.* **39**, S671-S684 (2006).
8. M. Mansuripur and A. R. Zakharian, "Maxwell's macroscopic equations, the energy-momentum postulates, and the Lorentz law of force," *Phys. Rev. E* **79**, 026608 (2009).
9. L. Landau, E. Lifshitz, *Electrodynamics of Continuous Media*, Pergamon, New York, 1960.
10. R. P. Feynman, R. B. Leighton, and M. Sands, *The Feynman Lectures on Physics*, Vol. II, Chap. 27, Addison-Wesley, Reading, Massachusetts (1964).

11. J. D. Jackson, *Classical Electrodynamics*, 2nd edition, Wiley, New York, 1975.
12. T. B. Hansen and A. D. Yaghjian, *Plane-Wave Theory of Time-Domain Fields: Near-Field Scanning Applications*, IEEE Press, New York (1999).
13. R. N. C. Pfeifer, T. A. Nieminen, N. R Heckenberg, and H. Rubinsztein-Dunlop, "Momentum of an electromagnetic wave in dielectric media," Rev. Mod. Phys. **79**, 1197-1216 (2007).
14. M. Mansuripur, "Radiation pressure and the linear momentum of the electromagnetic field," *Optics Express* **12**, 5375-5401 (2004).
15. M. Mansuripur, "Radiation pressure and the linear momentum of light in dispersive dielectric media," *Optics Express* **13**, 2245-2250 (2005).
16. M. Mansuripur, A. R. Zakharian, and J. V. Moloney, "Radiation pressure on a dielectric wedge," *Optics Express* **13**, 2064-74 (2005).
17. M. Mansuripur, "Angular momentum of circularly polarized light in dielectric media," *Optics Express* **13**, 5315-24 (2005).
18. A. R. Zakharian, M. Mansuripur, and J. V. Moloney, "Radiation pressure and the distribution of electromagnetic force in dielectric media," *Optics Express* **13**, 2321-36 (2005).
19. M. Mansuripur, "Radiation pressure and the distribution of electromagnetic force in dielectric media," *SPIE Proc.* **5930**, 0O-1:7 (2005).
20. M. Mansuripur, "Radiation pressure on submerged mirrors: Implications for the momentum of light in dielectric media," *Optics Express* **15**, 2677-82 (2007).
21. M. Mansuripur, "Momentum of the electromagnetic field in transparent dielectric media," *SPIE Proc.* **6644**, 664413 (2007).
22. M. Mansuripur, "Radiation pressure and the linear momentum of the electromagnetic field in magnetic media," *Optics Express* **15**, 13502-18 (2007).
23. M. Mansuripur, "Electromagnetic Stress Tensor in Ponderable Media," *Optics Express* **16**, 5193-98 (2008).
24. M. Mansuripur, "Electromagnetic force and torque in ponderable media," *Optics Express* **16**, 14821-35 (2008).
25. M. Mansuripur, "Generalized Lorentz law and the force of radiation on magnetic dielectrics," *SPIE Proc.* **7038,** 70381T (2008).
26. R. V. Jones and J. C. S. Richards, "The pressure of radiation in a refracting medium," *Proc. Roy. Soc. Lon. A* **221**, 480-498 (1954).
27. R. V. Jones and B. Leslie, "The measurement of optical radiation pressure in dispersive media," Proc. Roy. Soc. London, Series A, **360**, 347-363 (1978).
28. A. F. Gibson, M. F. Kimmitt, and A. C. Walker, "Photon drag in Germanium," Appl. Phys. Lett. **17**, 75-77 (1970).
29. Cs. Ferencz, O. E. Ferencz, D. Hamar, and J. Lichtenberger, *Whistler Phenomena: Short Impulse Propagation*, Kluwer Academic Publishing, Dordrecht, 2001.